\documentstyle[graphicx,float]{article}

\begin{document}
\title{Identical two-particle interferometry in diffraction gratings}
\author{Pedro Sancho \\ Delegaci\'on de AEMET en Castilla y
Le\'on  \\ Ori\'on 1, 47014, Valladolid, Spain}

\date{}
\maketitle
\begin{abstract}
We study diffraction and interference of indistinguishable
particles. We consider some examples where the wavefunctions and
detection probabilities  can be evaluated in an analytical way. The
diffraction pattern of a two-particle system shows notorious
differences for the cases of bosons, fermions and distinguishable
particles. In the example of near-field interferometry, the exchange
effects for two-fermion systems lead to the existence of planes at
which the probability of double detection is null. We also discuss
the relation with the approach to systems of identical particles
based on correlation functions. In particular, we shall see that
these functions reflect, as in noise interferometry, the underlying
periodic structure of the diffraction grating.
\end{abstract}

\section{Introduction}

Exchange effects, associated with the peculiar behaviour of
indistinguishable particles, have been extensively studied in the
literature. In recent times these studies have followed two
principal lines. On the one hand, the experimental measurement of
correlation functions has corroborated the existence of bunching and
antibunching effects. On the other hand, some interference phenomena
are strongly dependent on the distinguishable or undistinguishable
nature of the particles involved. In this paper we shall be mainly
concerned with this second line of research.

In the case of photons, following the seminal work of Hong-Ou-Mandel
(HOM)\cite{HOM}, the effects of the statistics of the particles have
been extensively studied for interferences originated by the
interaction in a beam splitter. Some authors have considered the
extension of that approach to the case of massive particles. In
\cite{Sil} (and references therein) the behaviour of fermions in Mach-Zehnder interferometers has been analyzed. The proposal presented in \cite{Gi1,Gi2}, is also interesting, where an
electronic HOM-type interferometer is used to detect entanglement.
In all the above arrangements the particles can only be detected in
the output arms of a beam splitter or interferometer. We analyse in
this paper whether exchange effects are also present in other
interferometric arrangements with a continuous spatial distribution.
The natural framework to discuss this possibility is that of
diffraction gratings, which generate continuous spatial patterns.

At this point we must signal the existence of other works by
Mandel's group where the spatial dependence of two-photon
interferences has been analysed \cite{Ma1,Ma2}: two photons
generated by down-conversion interfere at a beam splitter and detectors
are placed in the two output arms. Moving the detectors relative to

the beam splitter, a spatial two-photon interference pattern is
observed. At variance with our goal, these interferences are
generated at a beam splitter and not at a diffraction grating.

As a first approximation to the problem, we consider in this paper
some simple examples of diffraction and interference by gratings
which can be solved analytically. This way we can illustrate the
main ideas involved in the problem without addressing other more
technical questions. To be concrete, we shall analyse two
archetypical systems:

(i) {\it Diffraction by a single slit of a Gaussian wave packet.} This is
probably the simplest system where the effects can be analysed. We
shall evaluate the probability detection patterns showing sharp
differences between two-particle distributions of bosons, fermions and
distinguishable particles.

(ii) {\it Near-field interferometry with periodic gratings.} The best
known example of this type of arrangement is the Talbot Lau
interferometer \cite{Cla, Hac}, which has been used in a number of
interesting studies in matter wave interferometry \cite{Hac, Bre}.
In this case we shall find a novel characteristic associated with
the statistics of the system. For a pair of indistinguishable
fermions, there are planes with a null probability of double detection.

As signaled before, in the other principal line of study of
indistinguishable systems the focus is on the correlation functions.
This line originated in the seminal work of Hanbury Brown and Twiss
(HBT) \cite{HBT}, in which they counted joint detections (in two
separate detectors) of photons from different chaotic sources. The
correlations showed that the photons tend to arrive bunched in
groups. This behavior can be understood within a classical framework
if one introduces fluctuating phases. In contrast, the opposite
trend observed for fermions has no classical analogue. Then the only
framework able to completely describe bunching and
antibunching is the quantum one, where these effects can be
explained in an unified way in terms of constructive and destructive
interference. In the last years, the correlation functions of free
boson \cite{Yas,Fol, Jel} and fermions \cite{Hen, Rom, Jel} have
been experimentally obtained, corroborating the existence of
bunching and antibunching effects. It is natural to ask for the
behaviour of the correlation functions in interferometry and to see
if the presence of interference effects can modify the usual
picture. A well-suited technique for the position-sensitive counting
of particles when one is concerned with the spatial dependence of
the correlations is noise interferometry. This technique converts
spatial patterns into an interference signal. In particular, in
Refs. \cite{Fol,Rom} the existence of periodic quantum
correlations was shown between density fluctuations in an expanding atom cloud
when released from an optical trap. These spatial correlations
reflect the underlying ordering in the lattice trap. Noise
interferometry, is useful for identifying quantum phases of
ultracold atoms in periodic potentials. We shall evaluate the
correlation functions in the framework (ii) showing that, as in
noise interferometry, they reflect the periodic structure of the
underlying arrangement.

The plan of the paper is as follows. In Sect. 2 we present the basic
equations and derive some general properties of the interference of
two identical particles in a diffraction grating. The two
analytically solvable models considered in the paper are discussed
in Sects. 3 (diffraction) and 4 (near-field interferometry). The
last one is divided into two subsetions devoted, respectively, to
single- and multi-mode states. Section 5 deals with the behaviour of
correlation functions in diffraction gratings. Finally, in Sect. 6, we recapitulate on the principal results of the paper
and consider the possibility of testing them.

\section{General expressions}

The arrangement we consider here consists in a system of two
identical particles arriving on a diffraction grating. After the
diffraction grating we place detectors measuring the interference
pattern, which is obtained after many repetitions of the experiment.

We denote by $\psi _{\bf k} ({\bf x},t)$ and $\psi _{\bf p} ({\bf
y},t)$ the wavefunctions of the two particles, where ${\bf x}$ and
${\bf y}$ are the coordinates of the particles and ${\bf k}$ and
${\bf p}$ are their wavevectors or momenta. When the particles are
in single-mode states, they refer to the wavevectors or momenta of
these modes. On the other hand, if we are dealing with multi-mode
distributions, they represent the mean values of these distributions.
When the particles are identical the usual product wavefunction
$\psi _{\bf k} ({\bf x},t) \psi _{\bf p} ({\bf y},t)$, must be
replaced by
\begin{equation}
\frac{1}{\sqrt{2}} \left( \psi _{\bf k}({\bf x},t) \psi _{\bf p}
({\bf y},t) \pm \psi _{\bf p} ({\bf x},t) \psi _{\bf k} ({\bf y},t)
\right)
\end{equation}
In the double sign expressions the upper one holds for bosons and the lower one
for fermions. The probabilities associated with this wavefuntion are
\begin{eqnarray}
P({\bf x},{\bf y},t) =  \frac{1}{2}|\psi _{\bf k} ({\bf x},t)|^2|\psi _{\bf p} ({\bf y},t)|^2  + \frac{1}{2} |\psi _{\bf p}({\bf x},t)|^2|\psi _{\bf k} ({\bf y},t)|^2  \nonumber  \\
\pm Re(\psi _{\bf p}^*({\bf y},t) \psi _{\bf k}^*({\bf x},t) \psi
_{\bf k} ({\bf y},t) \psi _{\bf p} ({\bf x},t)) \label{eq:un}
\end{eqnarray}
We have an interference term that is not present for distinguishable
particles, according to the standard interpretation of exchange
effects as interference effects. From now on, in order to avoid any
possible confusion by the use of the interference concept in two
different ways, we restrict it to the effects induced by diffraction
gratings. The interference effects associated with the
indistinguishable character of the particles will be denoted as
exchange effects.

It must be remarked that the (anti)symmetrization procedure is only
physically required when there is a non-negligible overlapping
between the two wavefunctions. If not, the particles must be treated
as distinguishable ones. The important point is that once the
overlapping has taken place, due to the linearity of the evolution
equations, the (anti)symmetric form of the two-particle wavefunction
persists at subsequent times.

Several consequences directly emerge from the above expressions:

(a) For bosons in the same state (${\bf p}={\bf k}$), the probability
of simultaneous detection of two bosons at the same point is $
P({\bf x},{\bf x},t)=2|\psi _{\bf p} ({\bf x},t)|^4$. The
probability is proportional to the fourth power of the wavefunction
modulus. This behaviour was described in models of double ionization
by massive particles \cite{PLA} and double absorption of massive
particles \cite{AP}. At this point a word of caution is in order if
one tries to experimentally corroborate the above probability
distribution. The probability in Eq. (\ref{eq:un}) for different
points, ${\bf x} \neq {\bf y}$, can be experimentally measured using
two detectors placed at ${\bf x}$ and ${\bf y}$. In the case of
double (boson) detection at the same point ${\bf x}={\bf y}$, we can
only use one detector, which must be able to distinguish between
single- and double-detection events. This type of detector is
already available for photons \cite{But,Era}, but up to our
knowledge, not for massive particles.

(b) For fermions the wavefunction of the complete system can have
nodal points not present for factorizable ones. For example, if the
individual wavefunctions obey (up to a global phase) the relations
$\psi _{\bf k} ({\bf X},t) = \psi _{\bf p} ({\bf X},t)$ for a pair
of points ${\bf X}$ and ${\bf Y}$, these points are nodal ones of
the two-particle wavefunction (not present in the factorizable case
$\psi _{\bf k} ({\bf X},t) \psi _{\bf p} ({\bf Y},t)$ unless ${\bf
X}$ or ${\bf Y}$ are themselves nodal points). This property will
play a central role in the discussion of the detection distributions
in near-field arrangements.

(c) The set of points $({\bf X},{\bf Y},t)$ has another interesting
property (for bosons and fermions). The detection probabilities at
these points show a maximal deviation with respect to
distinguishable ones. A good measure to quantify how much the detection rate
is enhanced or diminished by the presence of the exchange effects is
given by the ratio of the detection rates for idistinguishable particles and
for the same state without exchange effects. It is defined at each point
and at a given time as
\begin{equation}
{\cal R}({\bf x},{\bf y},t)=\frac{P({\bf x},{\bf y},t)}{P_{NE}({\bf x},{\bf y},t)}
\end{equation}
where $P_{NE}$ denotes the probability distribution without interchange effects.
In the particular case of the set $({\bf X},{\bf Y},t)$ we obtain the maximum difference,
\begin{equation}
{\cal R}({\bf X},{\bf Y},t)= \frac{(1\pm 1)|\psi _{\bf k} ({\bf X},t)|^2|\psi _{\bf k} ({\bf Y},t)|^2}{|\psi _{\bf k} ({\bf X},t)|^2|\psi _{\bf k} ({\bf Y},t)|^2}=1\pm 1
\end{equation}
which is $2$ for bosons and $0$ for fermions. The rate of
simultaneous detections for bosons doubles that without exchange
effects, whereas that of fermions drops to zero. The resemblances
with the bunching and antibunching effects are clear, although in
this case the points ${\bf X}$ and ${\bf Y}$ (at a given time $t$)
do not need to be very close. The particles were close enough only
at a previous time, leading to the (anti)symmetrization of the
wavefunction, which persists at subsequent times.

In order to go beyond these general properties and to look for
other effects, we must consider particular arrangements and
evaluate the form of the wavefunctions in each situation. We do it in
the next two sections.

\section{Diffraction}

We consider in this section the diffraction of a pair of
indistinguishable particles by a slit. The rigorous analysis of the
problem leads to the numerical integration of Fresnel's functions
\cite{Fey, Ada}. A simpler approach to the problem is, following
Feynman \cite{Fey}, to replace the slit by a Gaussian slit. With
this approximation we have an analytically solvable problem. The
wave packet generated by this slit with {\it soft edges} has also a
Gaussian profile in the direction parallel to the slit. The movement
in the perpendicular direction is assumed to be unaffected. As usual
in the treatment of the problem, we consider a two-dimensional
problem, neglecting the {\it vertical} axis. Moreover, as the
movement in the perpendicular axis is almost unaffected, the relevant
physics is contained in the parallel axis and the problem reduces to
an one-dimensional one.

The simplest way to study multimode states is to assume that the
wavevector distribution of each particle is a Gaussian one
\cite{Ada}. Using the notation of Ref. \cite{Ada} the
one-dimensional Gaussian distribution is $f(k)=(4\pi )^{1/4} \sigma
^{-1/2} \exp (-(k-k_0)^2/2 \sigma ^2)$ with $\sigma $ the width of
the distribution and $k_0$ its central value. The wavefunction is
obtained by superposing a set of planes waves with that distribution
($(2\pi )^{-1}\int  dk f(k)\exp(i(kx-E_k t/\hbar))$ with $E_k= \hbar ^2 k^2/2m$)\cite{Ada}:
\begin{equation}
\psi (x,t)=C(t) \exp \left( \frac{-\sigma
^2(x-vt)^2+ik_0(2x-vt)+i\hbar \sigma ^4 x^2t/m}{\mu (t)} \right)
\label{eq:ttt}
\end{equation}
with $v=\hbar k_0/m$,
\begin{equation}
C(t)= \pi ^{-1/4} \left( \frac{1}{\sigma }+\frac{i\hbar \sigma t}{m}  \right)^{-1/2}
\end{equation}
and
\begin{equation}
\mu (t)= 2 \left( 1+ \frac{\hbar ^2 \sigma ^4 t^2}{m^2} \right)
\end{equation}
Note the absence with respect to the formula
in \cite{Ada} of a multiplicative coefficient $2$ in the last term
of the numerator in the exponential.

As required by our assumption of a Gaussian slit, the wavefunction
has the form (up to some phase terms) of a spatial Gaussian packet with
time-dependent width $\sigma ^{-2}+ (\hbar ^2 t^2 \sigma ^2 /m^2)$.

Using the expression of the wavefunction, it is immediate to evaluate
the probability distribution for two indistinguishable particles. We
assume both particles to be in states of the type (\ref{eq:ttt}), with
the same width and central wavevectors $k_0$ and $p_0$. The final
result is
\begin{equation}
P(x,y,t)/|C(t)|^4=\frac{1}{2}\left( e^A + e^B  \pm 2 e^{(A+B)/2}
\cos \left( \frac{2}{\mu (t)} (x-y)(k_0 - p_0)  \right) \right)
\label{eq:LLL}
\end{equation}
with
\begin{equation}
A=  -\frac{2 \sigma ^2}{\mu (t)} ( (x-v_k t)^2 +(y-v_p t)^2 )
\end{equation}
and $B$ given by a similar expression with the interchange of $v_k$ and $v_p$.

From the above expressions it is clear that the detection
probabilities for a pair of indistinguishable particles show
notorious differences with respect to those of distinguishable ones.
We illustrate this point numerically in Fig. 1, where we compare the
probability of simultaneous detection for bosons, fermions and
distinguishable particles. In order to simplify the presentation, we
fix one of the points, $x=x_0$, and evaluate the detection
probability for arbitrary $y$. We choose for $x_0$, at a fixed time
$t$, the value $x_0=v_k t$ which maximizes the detection probability
in the first exponential. The probability in the case of
distinguishable particles is $ P_{dis} (x,y,t)/|C(t)|^4 = e^A(x=x_0)
= \exp (-2 \sigma ^2 (y-v_p t)^2/\mu (t))$. Taking $\hbar t /m = 0.1$,
$k_0=1=-p_0$ and a width of the distribution $\sigma ^2 =0.125$ we have
the following distribution:
\begin{figure}[H]
\center
\includegraphics[width=5cm,height=5cm]{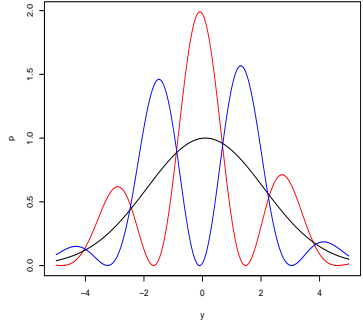}
\caption{In the vertical axis we represent, in arbitrary units,
$p=P(x_0,y,t)/|C(t)|^4$. In the horizontal axis we have the
coordinate $y$, also in arbitrary units, and the point
$0$ corresponding to $y=v_p t$. The red, black and blue curves correspond,
respectively, to bosons, distinguishable particles and fermions.}
\end{figure}
Several consequences emerge directly from the figure. As expected,
the distribution in the distinguishable case retains a Gaussian form
centred around the point $y=v_p t$. In contrast, for both fermions
and bosons, the mathematical form of the curves departs from the
Gaussian profile showing the typical form of an interference
pattern. The maximum probability for bosons is approximately at the
same position of the peak of the Gaussian for distinguishable
particles and doubles its value. At the same position we have a
null-probability point for fermions.

All the above development has been done in terms of the diffraction by a
slit. However, these expressions also describe two particles that interact
(there must be a strong overlapping between them) at a given moment
and then continue a free evolution. We can measure the positions of
the particles after the interaction, which are given by Eq.
(\ref{eq:LLL}). Note that in this case that expression is exact,
whereas for the description of the diffraction by the slit, it is
only an approximation.

\section{Near-field interferometry}

After considering an example of diffraction, we now move to a purely
interferometric arrangement, which illustrates other types of
effects associated with the interchange terms. In particular, we
focus on near-field interference in the eikonal approximation
\cite{Hac}, which can be tackled analytically (for the limitations
of the eikonal approximation see \cite{Nim}). One of the best known
examples of near-field arrangement is the Talbot Lau one \cite{Cla},
which has important applications in matter wave interferometry
\cite{Hac, Bre}. In order to emphasize on the main physical ideas of
the problem without excessive mathematical technicalities, in a
first step we shall only consider particles in single-mode states,
showing later that the main result persists (under adequate
restrictions) in the more realistic framework of multi-mode states.

\subsection{Single-mode states}

We consider a plane wave $\psi _o =\exp (ikz)$ which incides on a
grating placed in the plane perpendicular to the z-axis. If the grating
function is a periodic one, the transmission function is given by
\begin{equation}
{\cal T}(x)=\sum _n A_n \exp (2\pi inx/d)
\end{equation}
with $n$ an integer, $d$ the grating period and $A_n =|A_n|
\exp(i\xi _n)$ the coefficient of the term $n$. The wavefunction at
a distance $L$ behind the grating is \cite{Hac}:
\begin{equation}
\psi _k ^L(x)= e ^{ikL} \sum _n A_n \exp \left( \frac{2\pi i n x}{d}\right) \exp
\left( -\frac{i\pi n^2L\lambda _k}{d^2}\right)
\label{eq:lol}
\end{equation}
with $\lambda _k$ the wavelength of the particle and $x$ the
coordinate of the axis in which we measure the interferences.

We consider now the normalization of this wavefunction. From the
expression
\begin{equation}
|\psi _k^L (x)|^2 = \sum _{n,m} A_n^*A_m \exp \left( \frac{2\pi i
(m-n) x}{d}\right) \exp \left( -\frac{i\pi L(m^2-n^2)\lambda
_k}{d^2} \right)
\end{equation}
we see that $\int _{-\infty }^{\infty } |\psi _k^L (x)|^2 dx$ is not
bound. In effect, the integration of the constant terms  $A_0^*A_0,
A_1^* A_1, \cdots $ does not give a finite value. One possibility to
overcome this difficulty is to do the normalization in a relative
sense, ${\cal N}^{-1} \int _{-\infty }^{\infty } |\psi _k^L (x)|^2
dx=1 $, with ${\cal N}= (\sum _n |A_n|^2) \int _{-\infty }^{\infty }
dx$. An alternative form of normalization is to consider the
integration over the period $d$, $\frac{1}{d} \int _0^d dx |\psi
_k^L (x)|^2 = 1$. Taking into account that $\int _0^d dx \exp (2\pi
i n x/d)=d\delta _{n0}$, we have $\frac{1}{d} \int _0^d dx |\psi
_k^L (x)|^2 = \sum _n |A_n|^2 =A^2$, and the normalization condition
is obtained dividing by $A$. It is simple to see that the same
condition is reached using the relative normalization condition. The
normalized wavefunction after the diffraction arrangement is given
by $\psi _{k}^L(x)/A$, or equivalently by Eq. (\ref{eq:lol}) with
the replacement $A_i \rightarrow a_i=A_i/A$ ($a_i=|A_i/A|exp(i\xi
_n)$).

As anticipated in Sect. 2 let us consider the points $X$ where the
condition $\psi _k^L (X)= \psi _p^L (X) $ holds (up to a global
phase). In this example, the problem is stationary and we do not need
to include explicitly the time variable. We introduce the new
parameter
\begin{equation}
\phi _{kp}=\frac{\pi L}{d^2}(\lambda _k - \lambda _p )
\end{equation}
For all the initial wavevectors fulfilling the condition $\phi
_{kp}=2\pi n_* $, with $n_*$ an integer, we have that at all the
points $X$ contained in the plane defined by $z=L$, the wavefunctions
obey the relation
\begin{equation}
\psi _k^L (X)= e^{i(k-p)L} \psi _p^L (X)
\end{equation}
From now on, we restrict our considerations to fermions. The
antisymmetrized wavefunction at that plane becomes null:
\begin{eqnarray}
\frac{1}{\sqrt{2}} ( \psi _k^L (X) \psi _p^L (Y) - \psi _p^L (X) \psi _k^L (Y))
\rightarrow \nonumber \\
\frac{1}{\sqrt{2}} e^{i(k-p)L} ( \psi _p^L (X) \psi _p^L (Y) - \psi _p^L (X) \psi _p^L (Y)) =0
\end{eqnarray}
For all the initial wavevectors for which the relation $\phi
_{kp}=2\pi n_* $ holds, we have that in the plane $z=L$, there cannot
be two-fermion detections. Conversely, once the values of
$\lambda _k$ and $\lambda _p$ are fixed, there are always planes in which the
phenomenon can be observed. These planes are given by $z=L_{n}$,
with $L_{n}= 2n d^2/(\lambda _k - \lambda _p)$ and $n$ any
integer. All the points in these planes are nodal points of the
two-fermion wavefunction.

Note that this result does not forbid one-fermion detections at
these planes. In effect, the one-fermion detection probabilities are
given by the reduced detection probabilities, ${\cal P}(X)= \int
P(X,y)dy$. The integration on $y$ contains any $y$ placed on any
plane, not only the $Y's$ contained in the {\it nodal} plains. The
contributions of the points in the {\it nodal } plains are null,
$P(X,Y)=0$, but the contributions $P(X,y)$ with $y$ outside these
special planes can be different from zero. The sum of all these
contributions can be, in principle, different from zero leading to a
non-null one-fermion distribution at the {\it nodal} planes.

\subsection{Multi-mode states}

We move now to the more realistic case of non-monochromatic states.
We show that the result of the previous subsection is still valid
for narrow enough initial wavepackets.

If the momentum distribution of the wavepacket is $f(k)$, the wavefunction at
a distance $L$ behind the grating becomes
\begin{equation}
\psi _f ^L(x)= \sum _n A_n F_n \exp \left( \frac{2\pi i n x}{d}\right)
\label{eq:INM}
\end{equation}
with
\begin{eqnarray}
F_n= \int dk f(k) e ^{ikL} e^{(-i\pi n^2L\lambda _k /d^2)} \sim \nonumber \\
e ^{ik_0 L} e^{(-i\pi n^2L\lambda _{k_0} /d^2)} \int dk e^{-(k-k_0)^2/2\sigma ^2} e ^{i(k-k_0)L} e^{(-i\pi n^2L(k_0-k) /d^2kk_0)}
\end{eqnarray}
where we have adopted for the mode distribution a Gaussian curve
with the central wavevector $k_0$. In order to simplify the notation, we
have dropped the coefficients associated with its normalization.

The two imaginary exponentials inside the integral have a joint
argument $L(k-k_0)(1+(\pi n^2/d^2kk_0))$. When the condition
$d^2kk_0 \gg \pi n^2$ holds, the integral simplifies to $I(k_0)=\int
dk e^{-(k-k_0)^2/2\sigma ^2} e ^{i(k-k_0)L}$. Let us consider the
circumstances under which this approximation can be justified. Three
conditions must be fulfilled. (1) $n$ must be small. For large
values of $n$ the right-hand side of the inequality will be very
large and the term containing it cannot be neglected. Thus, the
summation in (\ref{eq:INM}) must be truncated at some value $n_0$.
This truncation is plainly justified because most of the behaviour
of the system can be described taking into account only the lower
terms in the summation. As a matter of fact, in the representative
experiment in \cite{Hac} the interference signal is essentially
determined by the $0$ and $\pm 1$ components only. (2) For values of
$k$ close to the central one, $k_0$, we must have $d^2 kk_0 \approx
d^2k_0^2 \gg \pi n_0^2$. This condition can be fulfilled by
restricting our considerations to this range of central wavevectors.
(3) As a consequence of (2), the distribution must be sharp enough
around $k_0$ in order the contribution of the elements with $|k| \ll
|k_0|$ to be negligible.

With this approximation Eq. (\ref{eq:INM}) becomes:
\begin{equation}
\psi _{k_0}^L(x)= \sum _{|n| \leq |n_0|} A_n I(k_0) e ^{ik_0 L} e^{(-i\pi n^2L\lambda _{k_0} /d^2)}   e^{(2\pi i n x/d)}
\end{equation}
Let us consider now another distribution with the same Gaussian form
(the same width), but centered around another value $p_0$. Then
$I(k_0)=I(p_0)$, and we are at the same position of the
previous subsection: when the relation $\phi _{k_0p_0}=2\pi n_* $
holds we have that in the plane $z=L$ the wavefunctions obey $\psi _{k_0}^L (X)= e^{i(k_0-p_0) L} \psi _{p_0}^L (X)$.

Within the range of validity of this approximation, the result
obtained in the previous subsection for a single-mode initial
wavefunction can be extended to its multi-mode counterpart.

\section{Correlation functions}

As signaled in the introduction, another fundamental line of
research in identical particles is based on correlation functions.
We analyse in this section whether the presence of the interference
device modifies the behaviour of the functions found in free
systems or optical lattices.

The correlation function is defined as
\begin{equation}
C(\eta )=\frac{1}{d}\int _{0}^{d} P(x,x+\eta )dx
\end{equation}
To calculate this expression we must evaluate the two-particle
probability distribution. In order to simplify the calculations, we
restrict our considerations to the case of particles described by
single-mode plane waves. As we have seen in the previous section, the result
obtained for a mono-chromatic particle can be extended to sharp
enough wave packets. We analyse the correlations in a fixed plane
$z=L$. The two detectors measuring the correlations must be placed
in that plane. Then it is not necessary to include the $L$ parameter
explicitly as a superscript of the wavefunction.

The modulus of the initial state is:
\begin{equation}
|\psi _k (x)|^2 = 1+ 2\sum _{n,m}^{m>n} |a_n a_m| \cos \left(
\frac{2\pi (m-n) x}{d}-\frac{\pi L(m^2-n^2)\lambda _k}{d^2}+ \xi _m
-\xi _n \right)
\end{equation}

In order to complete the expression for the probability, we must evaluate the crossed or interference term
\begin{eqnarray}
 Re(\psi _p^*( y) \psi _k^*(x) \psi _k(y)
\psi _p(x))=  Re  \sum _{n,m,r,s} a_n^* a_m^* a_r a_s
\times  \nonumber \\
\exp \left( \frac{2\pi i [(s-n) x +(r-m)y]}{d}\right) \exp \left(
-\frac{i\pi L[\lambda _p(s^2-m^2)+\lambda _k (r^2 - n^2)]}{d^2}
\right)
= \nonumber \\
\sum _{n,m,r,s} |a_n a_m a_r a_s| \times \\
\cos \left( \frac{2\pi  [(s-n) x +(r-m)y]}{d} -\frac{\pi L[\lambda _p(s^2-m^2)+ \lambda _k (r^2 - n^2)]}{d^2}+\xi _{r,s,n,m} \right) \nonumber
\end{eqnarray}
where $\xi _{r,s,n,m}=\xi _r + \xi _s -\xi _n -\xi _m$.
Adding the direct and crossed terms, we obtain for the probability
\begin{eqnarray}
P({\bf x},{\bf y})=1+ \nonumber \\
\sum _{m,n}^{m>n} |a_na_m|(\cos \varphi _{m,n}^k ({\bf x}) + \cos \varphi _{m,n}^p ({\bf x}) + \cos \varphi _{m,n}^k ({\bf y}) + \cos \varphi _{m,n}^p ({\bf y}) ) +  \nonumber \\
2 \sum _{n,m,r,s}^{m>n,r>s} |a_n a_m a_r a_s| (\cos \varphi _{m,n}^k ({\bf x}) \cos \varphi _{r,s}^p ({\bf y}) + \cos \varphi _{m,n}^p ({\bf x}) \cos \varphi _{r,s}^k ({\bf y})) \pm  \nonumber \\
\sum _{n,m,r,s} |a_n a_m a_r a_s| \times  \label{eq:cuac} \\
\cos \left( \frac{2\pi  [(s-n) x
+(r-m)y]}{d} -\frac{\pi L[\lambda _p(s^2-m^2)+ \lambda _k (r^2
- n^2)]}{d^2}+\xi _{r,s,n,m} \right) \nonumber
\end{eqnarray}
where $\varphi _{m,n}^k ( x)=\frac{2\pi  (m-n) x}{d}-\frac{\pi
L(m^2-n^2)\lambda _k}{d^2}+ \xi _m -\xi _n$. Now, the correlation function
can be obtained from Eq. (\ref{eq:cuac}). The calculation is simple
but lengthy. There are four types of contributions: (i) constant
terms, (ii) terms containing $\cos \varphi _{m,n}^k( x)$, (iii)
terms with $\cos \varphi _{m,n}^k( x) \cos \varphi _{s,r}^p(y)$ and
(iv) those contained in the crossed term. Let us evaluate them
separately.

(i) For constant terms we have, $\frac{1}{d} \int _0^d \; cte dx = cte$.

(ii) By direct calculation we have, $\frac{1}{d} \int _0^d \cos
\varphi _{m,n}^k(x) dx= 0 =\frac{1}{d} \int _0^d  \cos \varphi
_{m,n}^k(x+\eta ) dx$. The terms with this form do not contribute to
the correlation function.

(iii) Using standard trigonometric relations and integrals we have
that the integral $\frac{1}{d}\int _0^d \cos \varphi _{m,n}^k( x)
\cos \varphi _{r,s}^p(x+\eta ) dx$ equals $0$ when $m-n \neq \pm
(r-s)$ and
\begin{equation}
\frac{1}{2}\cos \left( \frac{2\pi (s-r)\eta }{d} + \phi _{mnsr}^{kp}
\right)+ \frac{1}{2}\cos \left( \frac{2\pi (s-r)\eta }{d} + \phi
_{mnsr}^{pk} \right)
\end{equation}
 otherwise, with $\phi _{mnsr}^{kp} =-\frac{\pi
L}{d^2} [(m^2-n^2)\lambda _k + (r^2-s^2)\lambda _p]+ \xi _m -\xi _n
-\xi _s + \xi _r$.

(iv) For the crossed term, we must distinguish when
$s-n+r-m$ is zero or different from zero. When it is not zero, we
have again the situation (ii) and its contribution is null. On the
other hand, when $s-n+r-m=0$, the integral of the cosine gives
\begin{equation}
\cos \left( \frac{2\pi (r-m)\eta }{d} +\Theta _{nmrs}^{kp} \right)
\end{equation}
with $\Theta _{nmrs}^{kp}=-\frac{\pi L}{d^2} [(r^2-n^2)\lambda _k + (s^2-m^2)\lambda _p]+ \xi _{r,s,n,m}$.

Combining all these expressions, we obtain the correlation function:
\begin{eqnarray}
C(\eta)=1+2\sum _{n,m,r,s}^{m>n,s>r} |a_ma_na_sa_r| \times \nonumber
\\ \left(\frac{1}{2}\cos \left( \frac{2\pi (s-r)\eta }{d} + \phi
_{mnsr}^{kp} \right) + \frac{1}{2}\cos \left( \frac{2\pi (s-r)\eta
}{d} + \phi _{mnsr}^{pk} \right) \right) \pm  \nonumber \\
 \sum _{n,m,r,s}^{s-n+r-m=0} |a_ma_na_sa_r| \cos \left( \frac{2\pi
(r-m)\eta }{d} + \Theta _{nmrs}^{kp} \right)
\end{eqnarray}
The correlation function depends on the parameters of the grating,
$a_i$ and $d$ and on the wavevectors of the incident particles
(through $\phi _{mnsr}^{kp}$ and $\Theta _{nmrs}^{kp}$ both
functions of $\lambda _k$ and $\lambda _p$). This behaviour is to be
compared with that of free particles, for which the correlation
function depends on the wavevectors in the form $C_{free}(\eta)=1\pm
\cos ((p-k)\eta)$ (for plane waves $\exp (ikx)$ and $\exp (ipy)$).
The spatial periodicity of $C_{free}$ is determined by $p-k$. For
interfering particles, however, the periodicity is only ruled by
$d/n$ (with $n$ any integer), a function of the parameter of the
grating $d$. In the correlation function we have all the periods
($d/n$) contained in the diffraction grating. The wavevectors only
determine (in conjunction with $d$ and the coefficients $\xi _i$) a
phase for each term. We conclude that, as in noise interferometry,
the correlation functions reflect the underlying structure of the
periodic grating.

A particularly clear illustration of the behaviour of the
correlation functions is obtained in the particular case that only
$a_1$ and $a_0$ are different from zero. Moreover, for the sake of
simplicity, we assume the two coefficients to be real. The
correlation function in this case is (using $a_1^2+a_0^2=1$)
\begin{equation}
C(\eta )=(1\pm 1)\pm 2a_0^2 a_1^2 (-1+\cos \phi _{kp}) + 2 a_0^2 a_1^2 (\pm 1+\cos \phi _{kp}) \cos \left( \frac{2\pi \eta}{d} \right)
\end{equation}
Now there is only one periodicity, $d$. The wavevectors only
determine (in conjunction with $a_o$ and $a_1$) the amplitude of the
oscillations. Moreover, the dependence is in the form
$\frac{1}{k}-\frac{1}{p}$, instead of $p-k$.

For short distances, $\eta \rightarrow 0$, we have $C_F(\eta ) \sim
\frac{C_0}{4} (\frac{2\pi }{d})^2\eta ^2$ ($C_0 = 4a_0^2 a_1^2
(1-\cos \phi _{kp})$)and $C_B(\eta )-C_B(0) \sim -a_0^2 a_1^2
(\frac{2\pi}{d})^2 (1+\cos \phi _{kp})\eta ^2$ that is the usual
dependence on $\eta ^2$, but with different coefficients.

\section{Discussion}

In this paper we have analysed the behaviour of exchange effects in
diffraction and interference. We have found the existence of some
distinctive characteristics of this type of interferometry:
modifications of the probability distributions with respect to those
of distinguishable particles, strong increase or decrease of the
double detection rates for some pairs of points (similar to bunching
or antibunching, but without need of closeness between the
particles) and existence of planes where the double detection of
fermions is forbidden. On the other hand, the correlation functions
show a behaviour similar to that found in noise interferometry,
reflecting the periodic structure of the diffraction grating.

Now briefly address the question of how the above distinctive
characteristics of continuous interferometry of identical particles
can be tested experimentally. As signaled before, the interchange
effects are present (and persist in the subsequent evolution of the
system) when there is a non-negligible overlapping between the
wavefunctions of the two particles at a given time. Then the key
step is to obtain a non-negligible overlapping of the two particles
at the time they reach the diffraction grating. This is equivalent
to a careful preparation of the times of arrival of the two
particles. In other words, the peaks of the two distributions must
reach the grating at the same time. Because of recent progress in
matter wave interferometry, interferometry of heavy molecules and
the existence of massive systems with high degrees of coherence,
these effects seem to be accessible to experimental scrutiny.
Conversely, the presence of the effects described in this paper
could be used as a test of the closeness of two particles under some specific preparation.

It must also be noted that because of the interaction of the beam
with the grating, there can be absorption and backscatter processes
which result in cases with zero or one particles arriving at the
detectors. We can overcome this difficulty by considering a
postselection process in which only events with two detections are
taken into account.

In this paper we have restricted our considerations to two simple
examples that can be treated analytically. Other types of
diffraction gratings must be studied in order to see if (the same or
different) exchange effects are also present.

{\bf Acknowledgments} We acknowledge partial support from MEC (CGL 2007-60797).

\end{document}